**Title :**

The main factors in student satisfaction with a campus environment: A mixed approach vs. a quantitative approach.

**First and last name :** Mohammed EDDAOU

**Function / Title:** Lecturer-Researcher

**Structure :**

Team member of the 'Number Theory, Cryptography and Computer Systems (TNCSI)', within the 'Arithmetic, Scientific Computing and Applications (ACSA)' research laboratory, Faculty of Sciences, Mohammed First University, Oujda, Morocco.

Lecturer-Researcher in Management Sciences at the Department of Management Sciences, Faculty of Legal, Economic and Social Sciences, Mohammed First University, Oujda, Morocco.

**Email:** mohammed.eddaou@ump.ac.ma.

**Abstract :**

University dropout rates in Morocco continue to increase, with approximately 49% of students leaving university before graduating, despite the successive reforms and measures taken to achieve Morocco's 2015–2030 strategic vision in the higher education sector "For a university of equity, quality and promotion", which raises questions about the state of knowledge on social inclusion at the university, capable of informing decision-making and the achievement of this strategic vision. While previous studies have used a quantitative approach with an exploratory purpose, to identify the main factors that affect the inclusion of students on university campuses. Knowledge that we consider insufficient to create general and regular knowledge, beyond the cases studied, on the exhaustiveness of these factors, no study has chosen a mixed approach (qualitative and quantitative) to create knowledge on the factors strengthening the attractiveness of the campus environment. Which brings us to our central question:

How does a mixed approach promote the creation of general and regular knowledge on the factors enabling the inclusion of students in the campus environment?

To answer our central question, we intend to discuss the external validity of future studies that will focus on the inclusiveness of the campus environment, mobilizing both a qualitative approach (interpretive descriptive research, phenomenology, ethnographic approach, case study and grounded theory) and a quantitative approach pursuing a descriptive purpose of the research (Structural equation modeling based on partial least squares 'PLS-SEM') and an explanatory one (Structural equation modeling based on covariance 'SEM-CB').

In a perspective scientific, this discussion will encourage researchers to choose a mixed approach, with a view to creating regular and general knowledge, beyond the cases studied by previous studies, thus participating in the construction of theories on the social inclusion of the university.

From a managerial perspective, this discussion will highlight the benefits of a mixed approach to framing future higher education reforms and decision-making, thereby enabling optimal allocation of resources, capable of creating an environment conducive to teaching and learning within universities, prioritizing efficiency over organizational isomorphism.

**Classification JEL :** I23, I30, H11, H83

**Introduction**

University dropout in Morocco continues to increase, approximately 49% of students leave university before graduating, despite the consecutive reforms and measures taken to achieve Morocco's strategic vision of 2015-2030 in the higher education sector "For a university of equity, quality and promotion". Therefore, a retention of 51% of students despite the consecutive reforms in front of a vision of a retention rate of 100%, finds from our point of view a reason in the theoretical underpinnings of the decisions taken by policy makers in the higher education sector. The knowledge developed on social inclusion at the university may have contributed to an organizational isomorphism at the level of the 12 Moroccan universities without the aim of efficiency, that is to say, achieving a desired retention rate. This raises questions about the state of knowledge on social inclusion at university, capable of informing decision-making and the realization of this strategic vision. While previous studies use a quantitative approach, but with an exploratory purpose, to identify the main factors that affect the inclusion of students on university campuses. Knowledge that we consider unsatisfactory for creating general and regular knowledge, beyond the cases studied, on the exhaustiveness of these factors, no study has chosen a mixed approach (qualitative and quantitative) to create knowledge on the factors strengthening the attractiveness of the campus environment. From our point of view, an exploratory quantitative approach does not create general and regular knowledge in the social sciences since the subjects' interpretative frameworks are constantly evolving and their behaviors present a certain heterogeneity. This brings us to our central research question:

How does a mixed approach promote the creation of general and regular knowledge on the factors enabling the inclusion of students in the campus environment?

To answer our central question, we intend to discuss the external validity of future studies that will focus on the inclusiveness of the campus environment, mobilizing both a qualitative approach (interpretive descriptive research, phenomenology, ethnographic approach, case study) and a quantitative approach pursuing a descriptive purpose of the research (Structural equation modeling based on partial least squares 'PLS-SEM') and an explanatory one (Structural equation modeling based on covariance 'SEM-CB'). And to provide methodological insight into our central question, we used the work of Tharenou, Donohue & Cooper (2007), Gavard-Perret et al. (2012), Corbière & Larivière (2014), Thiétart (2014) and Hair et al. (2017).

**1. An exploratory quantitative approach: Contributions and criticisms**

**Figure 1: Descriptive model (components and relationships)**

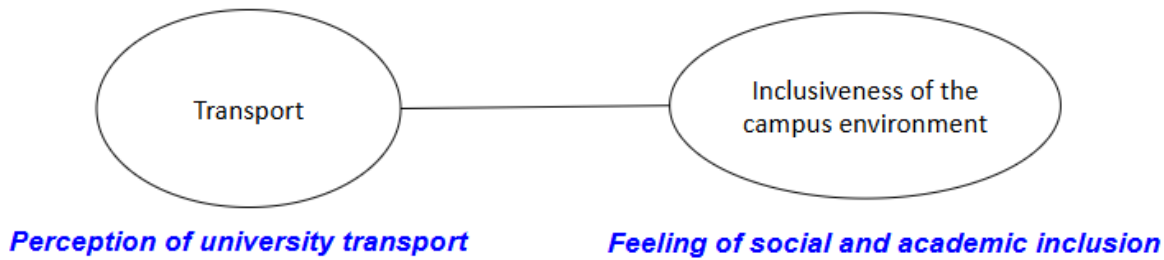

Source: Prepared by us[1]

The exploratory quantitative approach only allows us to detect components (satisfaction factors, inclusiveness of the campus environment) as well as the relationship between these different components (Thiétart, 2014, p.336-337). This association only reflects that student satisfaction factors and inclusiveness of the campus environment vary together without confirming a causal relationship (Thiétart, 2014, p.338).[2]. While increasing the inclusiveness of the campus environment requires addressing knowledge gaps among university decision-makers by identifying causes, on the basis of which they can act.

**Figure 2: Association and causality**

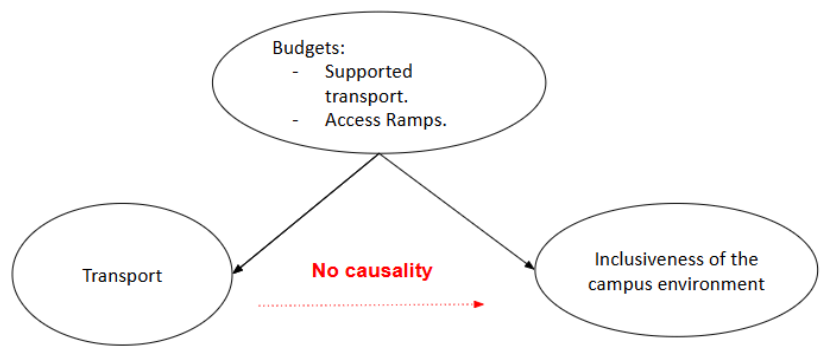

Source: Prepared by us

It may be that the joint variation between student satisfaction factors (an efficient university transportation system) and the inclusiveness of the campus environment (equitable access to campus buildings) is the consequence of a common cause such as the existence of a budget dedicated to subsidizing university transportation and the construction of access ramps. Under these conditions, university decision-makers will be able to act on the budget to improve the inclusiveness of the campus environment.

---

[1]Based on the work of Thiétart (2014, p.336-338).
[2]The author analyzes the relationships between two variables by identifying causal or association type relationships (Correlation or covariance).

**Figure 3: Normality of perception of university transport**

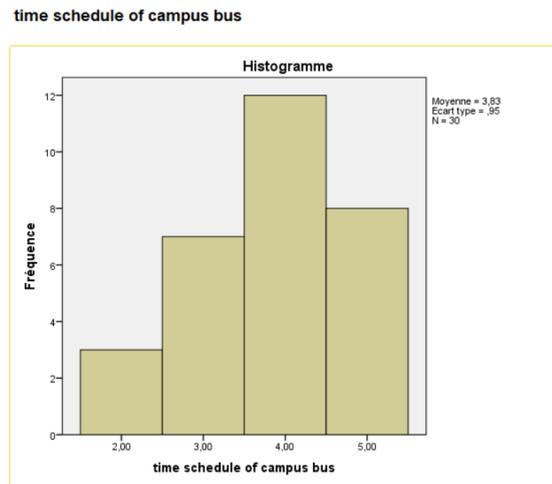

**Source: Prepared by us[3]**

Furthermore, and in an exploratory quantitative approach, the researcher is not concerned with the normality of the latent variables, which does not guarantee that the estimated parameters of a model that maximize the probability of observing the available data on the satisfaction factors and those of the inclusiveness of the campus environment, since the condition for applying a maximum likelihood estimation is multivariate normality (Hair et al., 2017).

Studies that adopt this approach do not highlight contextual regularities. They do not give us an idea about the external factors that maintained the relationship between these satisfaction factors as perceived by students (readability, social relations, quality of housing, facilities, extracurricular activities, accessibility, security, comfort, academic services, and transportation) and the equitable access of students to a campus environment (Corbière & Larivière, 2014, p.53).

---

[3]Using SPSS software.

**Figure 4: Contextual causal regularity**

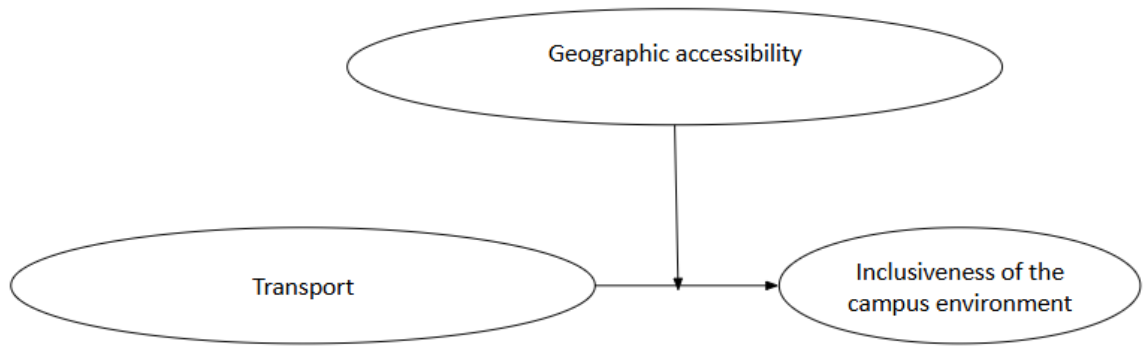

Source: Prepared by us[4]

For example, a student's perception of the importance of public transportation as a means of benefiting from equal access to a campus environment will only become regular knowledge if this knowledge is associated with a specific context, for example: Geographic accessibility (Thiétart, 2014, p.124). In this condition, the moderating variable "geographic accessibility" will accentuate or reduce the impact of university transportation on the inclusivity of the campus environment. For example, the level of satisfaction of students with the information provided on campus bus schedules would have an impact on their level of satisfaction with equal access and the sense of belonging to a campus environment, if the student's geographical accessibility to the latter poses a problem for them, and this additional resource, namely the availability of information, allows them equal access to this environment.

## 2. A mixed approach: Towards the creation of regular and general knowledge

The strength of the mixed approach is that it will allow us to design and validate a consistent explanatory model relating to the inclusiveness of the campus environment. It will contribute to the formulation of a theory on the inclusivity of the campus environment by providing a satisfactory justification for the question "why"? (Tharenou, Donohue & Cooper, 2007, p.7-8). Therefore, higher education policy makers and practitioners will make informed decisions based on a consistent and validated explanatory model.

The mixed approach will allow future research, firstly, to use a qualitative approach. This approach would promote the design of an explanatory model identifying the causes whose decisions will lead to improving the inclusiveness of the campus environment, the intermediate variables forming the impact mechanisms, and the moderating variables ensuring contextual regularity. And secondly, to use a quantitative approach, in order to confirm the causal regularity in the sample and successfully generalize this causal relationship to the parent population (university campuses in Morocco).

---

[4]based on the work of Corbière & Larivière (2014, p.53)

**2.1 Qualitative study**

**2.1.1 Interpretive descriptive research**

**Figure 5: Describe the components of the phenomenon and the plausible relationships**

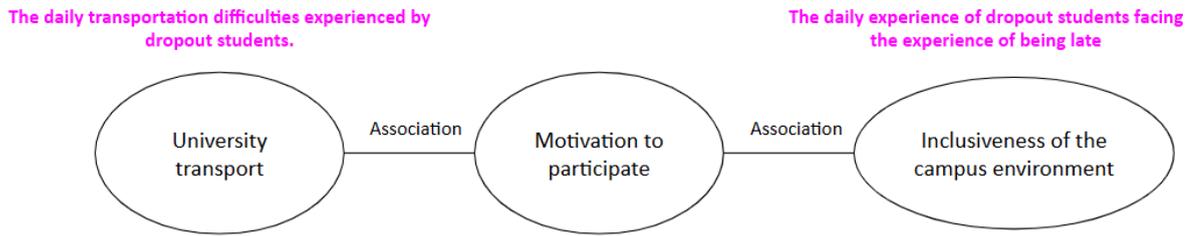

**Source: Prepared by us**[5]

The usefulness of interpretative descriptive research in explaining the inclusivity of the campus environment lies in producing descriptions of the psychosocial needs of dropout students following a past experience of the campus environment (Corbière & Larivière, 2014, p.5). It allows us to identify the components of students' needs in a campus environment, as well as the plausible relationships between these different components (Descriptive model required by an explanatory model) (Corbière & Larivière, 2014, p.8). This research will allow us to adopt the perspective of the students themselves and to recognize that there may be several ways of experiencing the campus environment (Corbière & Larivière, 2014, p.11) and to come out with the emerging needs that are repeated. If dropout students felt frustrated by always being late to class, and they expressed a need for justice with their classmates by wanting equitable access to resources, such as transportation. The availability of university transportation for excluded students could therefore promote the inclusiveness of the campus environment (Equitable access to class sessions). In this perspective, we will avoid imposing a pre-existing theoretical framework in favor of descriptions and interpretations emerging from the field.

---

[5]based on the work of Corbière & Larivière (2014, p.5, 8, 11).

**Figure 6: Reliable and valid measurement indicators**

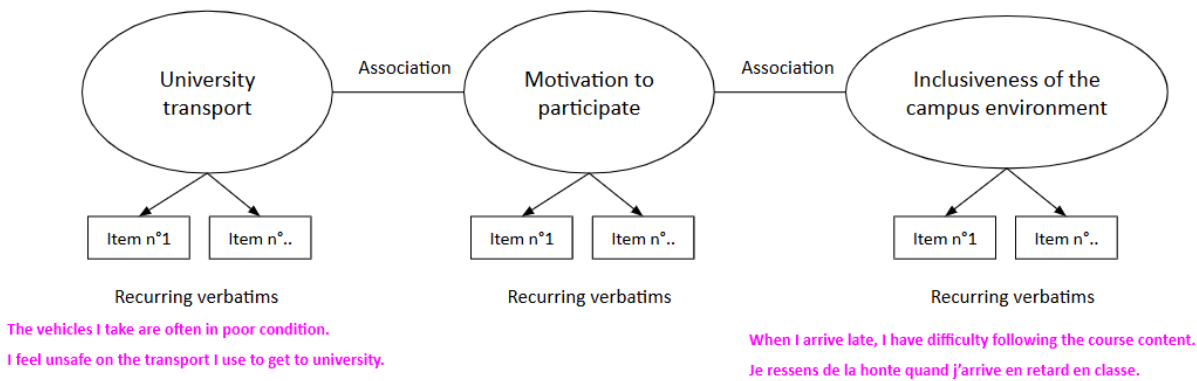

Source: Prepared by us[6]

Furthermore, the usefulness of interpretative descriptive research in explaining the inclusivity of the campus environment is that it will help us identify the psychosocial needs of dropout students following the shared meanings of the feelings experienced during the experience of an exclusive campus environment. These can become an intersubjective objective reality (Gavard-Perret et al., 2012, p. 38), and promotes the creation of reliable and valid measurement models required for the design of a consistent explanatory model on the inclusivity of the campus environment, so that the results of its empirical study can be deemed reproducible, generalizable and cumulative (Thiétart, 2014, p. 330).

**Figure 7: The extended coverage of the dimensions of the construct "University transport"**

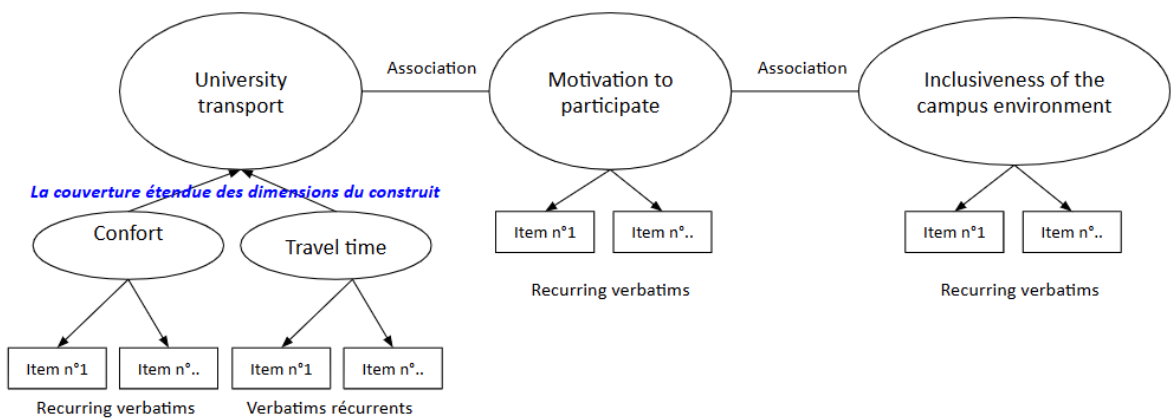

Source: Prepared by us

A multi-level categorization of the components of student needs resulting from the lived experience of the campus environment, makes it possible to recover the information required to have a causal relationship between the different components (Diamantopoulos & Winklhofer, 2001), (Corbière & Larivière, 2014, p.13) and (Hair et al., 2017). University transportation that creates suffering for students in terms of comfort and travel time could create a demotivation for students to participate in a campus environment. We believe that covering the different dimensions of a construct representing an explanatory variable helps capture the information needed to apply a significant impact on another construct representing a variable to be explained. In this perspective, the conceptual model designed will provide a sufficient answer to the question: why is there an inclusive campus environment?, and build a theory of social inclusion at the university capable of informing the policies of decision-makers and practitioners in higher education.

---

[6]based on the work of Gavard-Perret et al. (2012, p.38) and Thiétart (2014, p.330).

### 2.1.2 The grounded theory

**Figure 8: Identification of causal mechanisms**

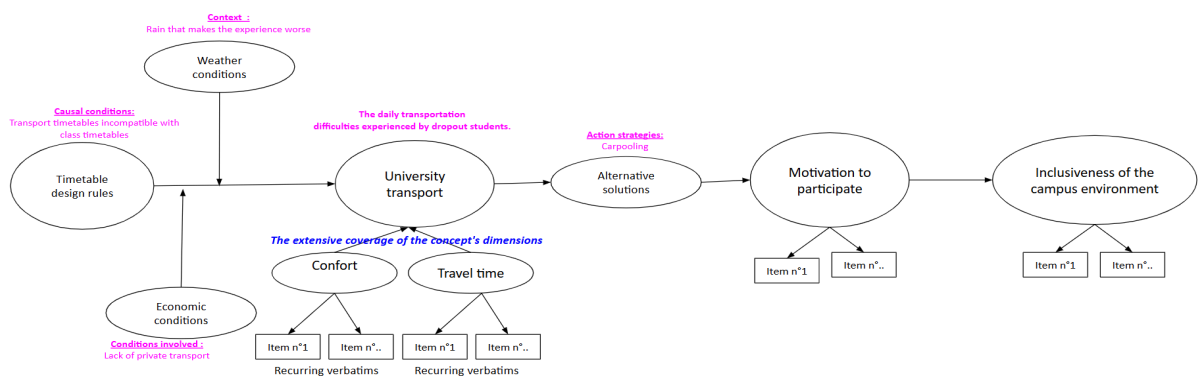

**Source: Prepared by us**[7]

For its part, the grounded theory aims to examine in depth the processes underlying the needs of students (justice of access to the course on time, etc.) having experienced a campus environment (A class of courses) (Corbière & Larivière, 2014, p.97-107). It will allow us to **explore the social structures** organizing the life of students in a campus environment (the rules of design of the timetable, the structures of transport and accessibility, ...). If the modules are scheduled to be taught separately, at the rate of one module per day, and in the face of the unavailability of university transport, the student would be demotivated to attend the courses, and the inequality of access to the latter would be the daily experience of this excluded student. Furthermore, and according to Thiétart, 2014 (p.207-208), it will allow us to identify :

- the causal conditions (Why the negative perception of transport (Phenomenon)?): Incompatible transport schedules with the students' course schedules,

---

[7]based on the works of Corbière & Larivière, 2014 (p.97-107) and Thiétart, 2014 (p.207-208).

- the context where the negative perception (When, where, how? Rain that worsens the experience),
- the strategies of actions and interactions engaged by the students (carpooling as an alternative solution),
- the intervening conditions (factors facilitating or constraining the negative perception of transport): the economic conditions of the students that do not allow them to use private means of transport
- and the consequences of the strategies for action and interaction adopted by students : A demotivation to participate in campus life and a feeling of social and academic exclusion.

### 2.1.3 Ethnographic approach

**Figure 9: Contextual causal regularity**

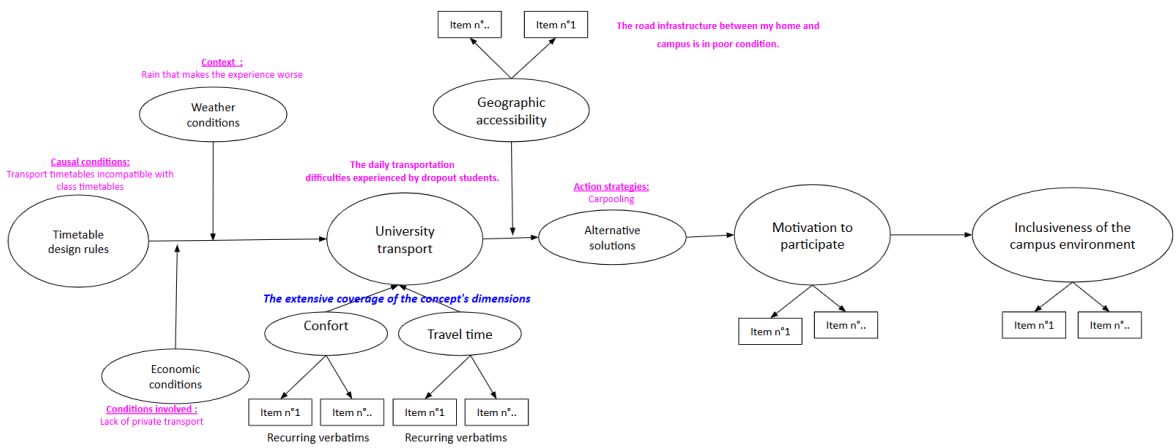

**Source: Prepared by us[8]**

The ethnographic approach allows the researcher to be present in a campus environment for a long period of time, which allows him to describe in depth the satisfaction factors of a campus environment studied as it is experienced and perceived on a daily basis (Corbière & Larivière, 2014, p.53). It analyzes external factors that influenced the emotions of students (the frustration of coming late to class) felt during their experience of a campus environment. For example, the geographical accessibility of students (the distance between the students' place of residence and the campus, the isolation of certain poorly served rural areas, degraded or insufficient road infrastructure) could be considered among the conditions that accentuate the effect of the availability of university transport on the inclusivity of the campus environment. In this case, university transport would promote the inclusivity of the campus environment in the case where there is a problem of geographical accessibility of students. This approach therefore allows us to identify contextual causal regularity and to contribute to the consistency of our causal model on the inclusiveness of a campus environment.

---
[8]based on the work of Corbière & Larivière (2014, p.53).

### 2.1.4 Phenomenological approach

**Figure 10: Reliability and validity of measurement models in a SEM**

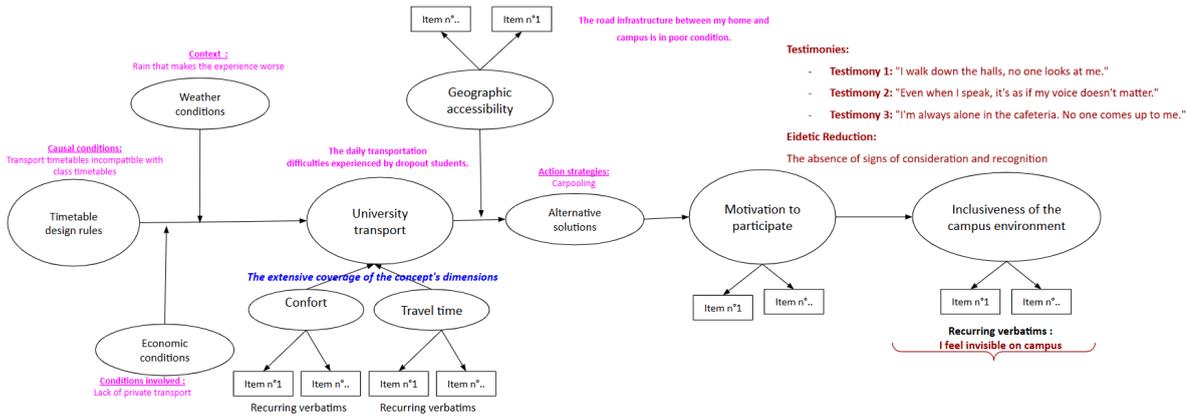

**Source: Prepared by us[9]**

For its part, the phenomenological approach will also contribute to the formulation of reliable and valid measurement models in structural equation modeling, since it participates in the emergence of a pure and universal meaning (essence) of the experience of students in a campus environment. An essence that presents the common discourses coming from the discourses of dropout students (Corbière & Larivière, 2014, p.29-32). It thus contributes to the identification of an observable, measurable and reproducible intersubjective objective reality.

### 2.1.5 The case study

**Figure n°11: Holistic understanding**

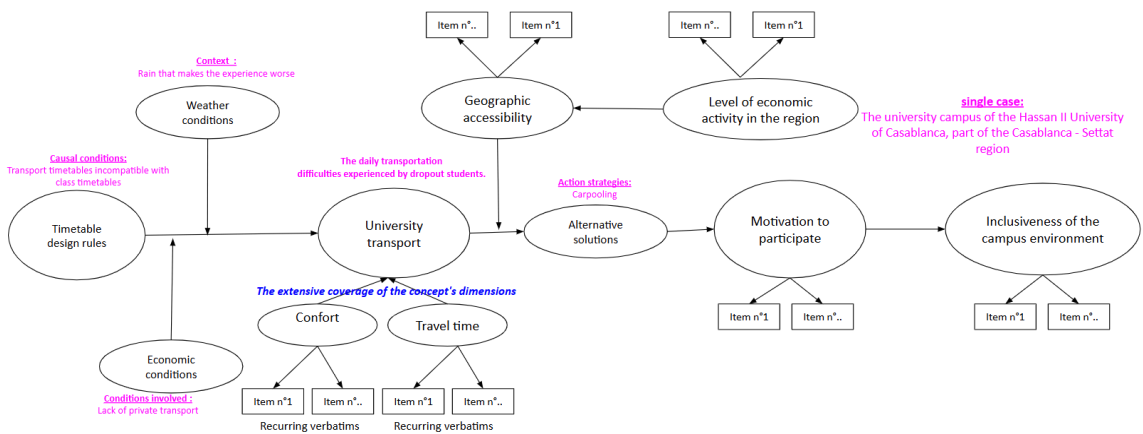

---

[9]based on the work of Corbière & Larivière (2014, p.29-32).

**Source : Established by us[10]**

The case study, as a qualitative research approach, could further feed our explanatory model, and this through the generation of a holistic understanding, and without manipulation by the researcher, on the why and how, an inclusiveness of a university campus environment is established, through a study of this particular phenomenon (the case of a university campus with an inclusive environment) in its natural context and through the use of a triangulation of the collected data (Corbière & Larivière, 2014, p.73-74). And we can for example, be interested in a single case: The university campus of the Hassan II University of Casablanca, part of the Casablanca - Settat region. In this case, the high level of economic activity favors the geographical accessibility of students to the campus environment.

## 2.2 Quantitative study

### 2.2.1 An exploratory quantitative study

---

[10]based on the work of Corbière and Larivière (2014, p.73-74).

**Figure No. 12: The Contribution of Causal Mechanisms (SEM-PLS)**

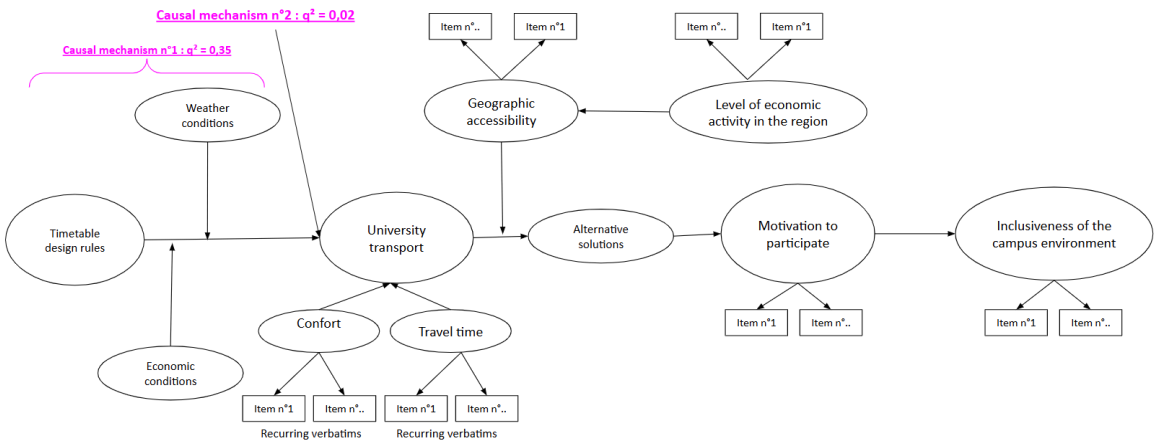

**Source: Prepared by us[11]**

According to Hair et al. (2017), an exploratory quantitative study that will focus on the factors that promote the inclusivity of the campus environment will reveal how the variables are related in order to reduce the variables into composite variables.

**Figure 12: Evaluation of the structural model (SEM-PLS)**

- Coefficients of determination ($R^2$)
- Predictive relevance ($Q^2$)
- Size and significance of path coefficients
- Effect size $f^2$
- Effect sizes $q^2$

**Source: Prepared by us[12]**

It will allow us to measure the size of the effect of each of the explanatory variables in the causal model, with the aim of verifying the contribution of each causal mechanism in the variation of the explanatory variable "the inclusiveness of the campus environment".

**2.2.2 An explanatory and predictive study**

---

[11]based on the work Hair et al. (2017)
[12] based on the work Hair et al. (2017) and the presentation of (EDDAOU, 2025, February 18; EDDAOU, 2025, February 25)

Figure 13: Confirmation of causal mechanisms (SEM-CB)

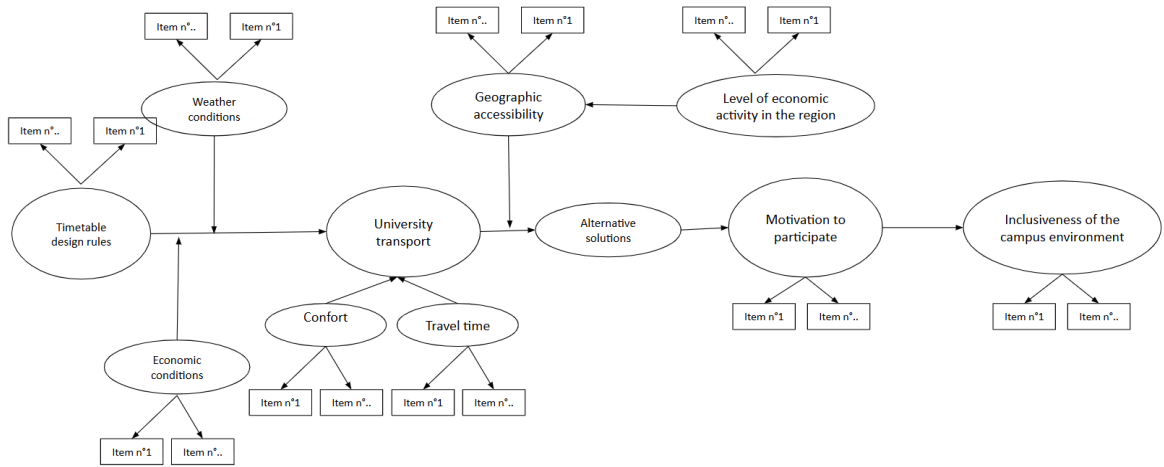

Source: Prepared by us[13]

In a quantitative approach where the study is explanatory and predictive (Thiétart, 2014, p.17) and (N'da, 2015, p.26), the researcher aims to discover stable causal regularities between the satisfaction factors (readability, social relations, quality of housing, facilities, extracurricular activities, accessibility, security, comfort, academic services, and transport)[14] and the inclusivity of a campus environment. When, for example, positive perceptions of transportation increase, there is a high probability that the inclusivity of a campus environment will also increase. The aim is to find consistent relationships in cross-sectional data between satisfaction factors and the inclusivity of a campus environment, capable of informing decision-making by higher education policy makers and practitioners.

Figure 14: Evaluation of the structural model (SEM-CB) and statistical inference

- Multivariate normality.
- Global fit: Based on the difference between the empirical covariance matrix and the implicit (theoretical) covariance matrix of the model.
- Local fit.
- Significance test of internal weights.

Source: Prepared by us[15]

---

[13] based on the work of Thiétart, 2014 (p.17) and N'da, 2015 (p.26).
[14] according to Sedaghatnia et al. (2015).
[15] based on the work Hair et al. (2017) and the presentation of (EDDAOU, 2025, February 18; EDDAOU, 2025, February 25)

Overall fit tests, such as the chi-square test, will allow us to confirm that the small difference observed in the sample between the empirical covariance matrix and the theoretical covariance matrix implicit in the model is significant. The chi-square test compares the data sample with the fitted covariance matrices in the model. Ideal value of the significance threshold: p-value > 0.05:

H0: The estimated model fits the data well, meaning that the observed and fitted covariance matrices do not differ significantly.

### 2.2.3 Bootstrapping

According to Hair et al. (2017), bootstrapping plays an important role in evaluating the coefficients of the structural paths of the constructed structural equation modeling, which will contribute to the generalization of the causal model created on the inclusiveness of a campus environment.

### Conclusion

Faced with a constantly increasing university dropout rate in Morocco, it is necessary to question the state of knowledge on social inclusion at university, capable of informing decision-making and the realization of this strategic vision. While previous studies have used a quantitative approach, but with an exploratory purpose, to identify the main factors that affect the inclusion of students on university campuses. This knowledge, we consider unsatisfactory for creating general and regular knowledge, beyond the cases studied, on the exhaustiveness of these factors, no study has chosen a mixed approach (qualitative and quantitative) to create knowledge on the factors that strengthen the attractiveness of the campus environment. This brings us to our central question:

How does a mixed approach promote the creation of general and regular knowledge on the factors enabling the inclusion of students in the campus environment?

As a methodological contribution, we have presented in this research the advantages of a mixed approach in explaining factors that enhance the attractiveness of the campus environment, by creating regular and general knowledge that answers the question of (why ?). A mixed-method approach will be based on a qualitative approach to data collection and analysis. This approach allows for in-depth descriptions of the factors of satisfaction and the inclusiveness of the campus environment as they manifest themselves in the natural campus environment, and constructed from the students' point of view, allows for an understanding of these phenomena and the grasp of their components and the relationships between them. A quantitative approach, subsequently, will be used to verify the causal regularity and its generalization. And we believe that this general and regular knowledge created on the basis of a mixed-method approach is capable of informing decision-making and the realization of this strategic vision.

As a research limitation, it remains to empirically verify the superiority of a mixed approach over an exploratory quantitative approach and its relevance to decision-making and the achievement of this strategic vision.

As a perspective for this research, a mixed approach promoting a consistent, valid and general explanatory model essential for decision-making requires human and financial resources. In this sense, this work suggests the realization of a collective research project. This is a deliverable that has a start date and an end date allowing the mobilization of the different relevant qualitative and quantitative methods.